\documentclass[journal,onecolumn]{IEEEtran}

\usepackage[pdfstartview=XYZ,
bookmarks=true,
colorlinks=true,
linkcolor=blue,
urlcolor=blue,
citecolor=blue,
pdftex,
bookmarks=true,
linktocpage=true, 
hyperindex=true
]{hyperref}
\usepackage{lipsum}
\usepackage{changepage}
\usepackage{orcidlink}
\usepackage{tgpagella}
\usepackage{tabularx}
\usepackage{booktabs}
\usepackage{cleveref}
\usepackage{tikz}
\usepackage{svg}
\usepackage{etoolbox}
\usepackage[none]{hyphenat}
\usepackage{microtype}
\usepackage{hyperref}
\patchcmd{\section}{\centering}{}{}{}
\graphicspath{ {./figures/} }

\newcommand*\circled[1]{\tikz[baseline=(char.base)]{
            \node[shape=circle,color=black, draw,inner sep=0.8pt, minimum size=2pt] (char) {#1};}}
\newcommand{\rpoint}[1]{\circled{{\fontfamily{pcr}\selectfont\footnotesize{#1}}}}
\definecolor{tab_blue}{RGB}{31, 119, 180} 
\definecolor{tab_orange}{RGB}{255, 127, 14} 
\definecolor{tab_green}{RGB}{44, 160, 44} 
\definecolor{tab_red}{RGB}{214, 39, 40} 
\definecolor{tab_purple}{RGB}{148, 103, 189} 

\newcolumntype{C}{>{\centering\arraybackslash}X} 
\newcolumntype{L}{>{\raggedright\arraybackslash}X} 

\ifCLASSINFOpdf
\else
\fi
\begin{document}
%
\title{\flushleft\LARGE\textbf{A Homomorphic Encryption Framework for Privacy-Preserving Spiking Neural Networks}}
%
%
%

\author{
    \IEEEauthorblockN{\flushleft\normalsize\textbf{Farzad Nikfam$^{1}$\orcidlink{0000-0003-1666-2677}, Raffaele Casaburi$^{1}$, Alberto Marchisio$^{2}$\orcidlink{0000-0002-0689-4776}, Maurizio Martina$^{1}$\orcidlink{0000-0002-3069-0319} and Muhammad Shafique$^{2}$}\orcidlink{0000-0002-2607-8135}}
    \newline\newline
    \footnotesize\IEEEauthorblockA{$^{1}$   Department of Electrical, Electronics and Telecommunication Engineering, Politecnico di Torino, 10129 Torino TO, Italy; farzad.nikfam@polito.it (F.N.); s266053@studenti.polito.it (R.C.); maurizio.martina@polito.it (M.M.)}
    \newline
    \footnotesize\IEEEauthorblockA{$^{2}$   eBrain Lab, Division of Engineering, New York University Abu Dhabi, UAE; alberto.marchisio@nyu.edu (A.M.); muhammad.shafique@nyu.edu (M.S.)}
}

\maketitle


\normalsize\noindent\textbf{Abstract:} Machine learning (ML) is widely used today, especially through deep neural networks (DNNs); however, increasing computational load and resource requirements have led to cloud-based solutions. To address this problem, a new generation of networks has emerged called spiking neural networks (SNNs), which mimic the behavior of the human brain to improve efficiency and reduce energy consumption. These networks often process large amounts of sensitive information, such as confidential data, and thus privacy issues arise. Homomorphic encryption (HE) offers a solution, allowing calculations to be performed on encrypted data without decrypting them. This research compares traditional DNNs and SNNs using the Brakerski/Fan-Vercauteren (BFV) encryption scheme. The LeNet-5 and AlexNet models, widely-used convolutional architectures, are used for both DNN and SNN models based on their respective architectures, and the networks are trained and compared using the FashionMNIST dataset. The results show that SNNs using HE achieve up to 40\% higher accuracy than DNNs for low values of the plaintext modulus $t$, although their execution time is longer due to their time-coding nature with multiple time steps.

\vspace{5mm}


%
\IEEEpeerreviewmaketitle

\par\noindent\rule{\textwidth}{0.4pt}

\section{\upshape\textbf{Introduction}}
Machine learning (ML) has witnessed significant development in recent years, finding diverse applications in various sectors such as robotics, automotive, smart industries, economics, medicine, and security~\cite{DBLP:journals/access/CapraBMMMS20, DBLP:conf/vts/DaveMHGSAS22, DBLP:conf/iccad/0001MPH21}. Several models based on the structure of the human brain have been implemented~\cite{DBLP:journals/spm/SimeoneRGEDDH19}, including the widely used deep neural networks (DNNs)~\cite{DBLP:journals/nn/Schmidhuber15, DBLP:journals/cacm/KrizhevskySH17} and spiking neural networks (SNNs)~\cite{von_Kugelgen_2017}, which emulate the functioning of neurons relatively better than DNNs~\cite{10.3389/fnins.2015.00491}. {These models require large amounts of data to be trained and reach high accuracy. However, if such data are collected from users' private information, such as personal images, interests, web searches, and clinical records, the DNN deployment toolchain will access sensitive information that could be mishandled}~\cite{DBLP:conf/mmsec/BarniOP06}{. Moreover, the large computational load and resource requirements for training DNNs have led to outsourcing the computations on the cloud, where untrusted agents may undermine the algorithms' confidentiality and intellectual property of the service provider. Note that encrypting the data transmission in the communication from client to server using common techniques such as advanced encryption standard (AES) would not solve the issues, because untrusted agents on the server side have full access to the sensitive data and DNN model.} Among privacy-preserving methods, homomorphic encryption (HE) employs polynomial encryption to encrypt input data, perform computations, and decrypt the output. \linebreak {Because the computations are conducted in the encrypted (ciphertext) domain, the ML algorithm and data remain confidential as long as the decryption key is unknown to the adversary agents.} However, common HE-based methods focus on traditional DNNs, and studying the impact and potential of encryption techniques for SNNs is still unexplored.

In this work, we deploy the Brakerski/Fan-Vercauteren (BFV) HE scheme~\cite{DBLP:journals/iacr/FanV12} for SNNs, and compare with its application to DNN architectures~\cite{DBLP:journals/pieee/LeCunBBH98}. From the experimental results, we observed that the SNN models working on encrypted data yield better results than traditional DNN models, despite the increased computational time due to the intrinsic latency of SNNs that simulate human neurons.

\begin{figure}[h]
    \centering
    \includegraphics[width=.7\linewidth]{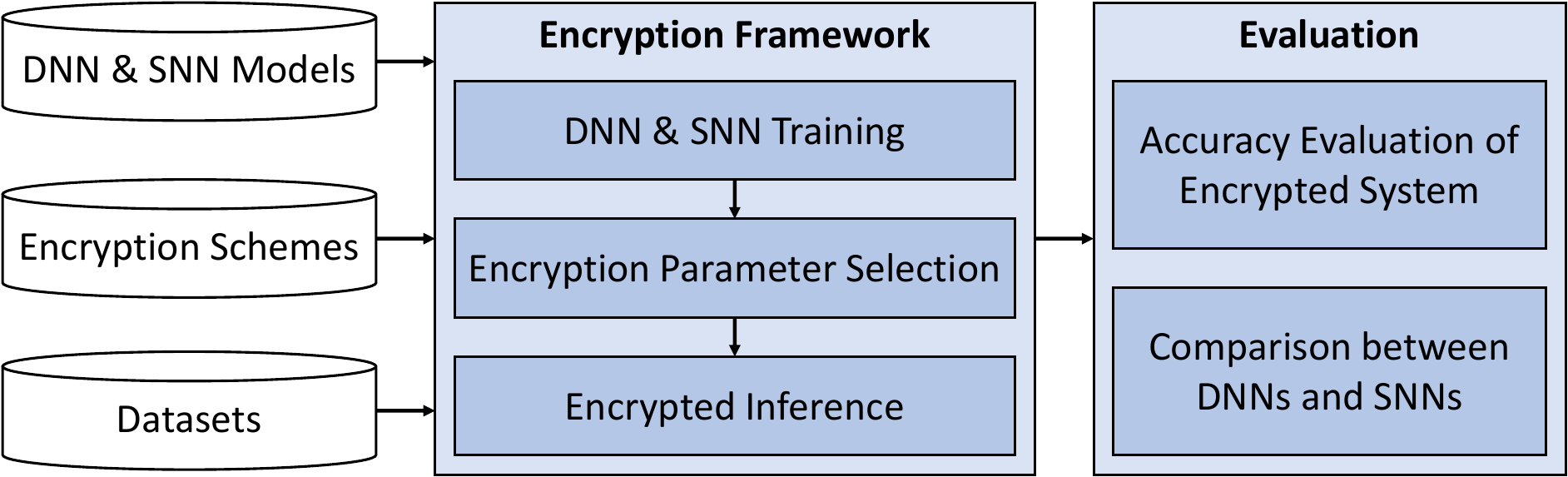}
    \caption{Overview of our novel contributions.}
    \label{fig:novel_contrib}
\end{figure}

Our novel contributions are summarized as follows (see an overview in Figure~\ref{fig:novel_contrib}):

\begin{itemize}
    \item We design an encryption framework based on the BVF HE scheme that can execute privacy-preserving DNNs and SNNs ({Section~\ref{sec:proposed_encryption_framewrok}}).
    \item The encryption parameters are properly selected to obtain good tradeoffs between security and computational efficiency ({Section~\ref{subsec:HE_parameters}}).
    \item We implement the encryption framework, evaluate the accuracy of encrypted models, and compare the results between DNNs and SNNs. We observe that the SNNs achieve up to 40\% higher accuracy than DNNs for low values of the plaintext modulus $t$ ({Section~\ref{sec:results}}).
\end{itemize}

Paper organization: Section~\ref{sec:background} contains the background information of the methods and algorithms used in this work, which are DNNs, SNNs, and HE, with a particular focus on the BFV scheme. Section~\ref{sec:proposed_encryption_framewrok} discusses the proposed encryption framework for DNNs and SNNs and describes our methodology for selecting the encryption parameters. \linebreak Section~\ref{sec:results} reports the experimental results and a discussion on the comparison between DNNs and SNNs when using HE. Section~\ref{sec:conclusion} concludes the paper.

\section{\upshape\textbf{Background}}
\label{sec:background}

\subsection{Deep Neural Networks and Convolutional Neural Networks}
\begin{figure}[h]
    \centering
    \includegraphics[width=\linewidth]{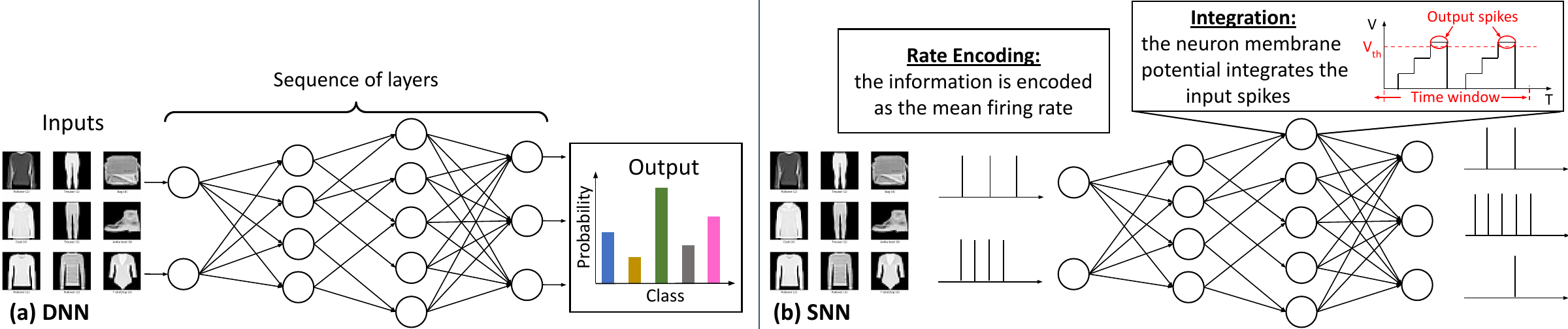}
    \caption{Overview of \textbf{(a)} the functionality of a DNN and \textbf{(b)} the functionality of an SNN.}
    \label{fig:SNN}
\end{figure}

DNNs, whose functionality is shown in Figure~\ref{fig:SNN}a, are a class of artificial neural networks composed of multiple layers of interconnected nodes called neurons.\linebreak These networks are designed to mimic the structure and functioning of the human brain. DNNs are characterized by depth, referring to the many hidden layers between the input and output. This depth allows DNNs to learn complex patterns and representations from data, enabling them to solve intricate problems in fields such as image and speech recognition, natural language processing, and more. 

Convolutional neural networks (CNNs)~\cite{DBLP:journals/ijon/JohnsonLC00} are a specialized type of DNN designed to efficiently process grid-like data, such as images or time series. CNNs apply filters to input data, capturing local patterns and features. This allows CNNs to extract hierarchical representations from visual data, enabling object detection, image classification, and image generation tasks. CNNs have revolutionized the field of computer vision and have been widely adopted in various applications, including autonomous driving, medical imaging, and facial recognition.

\subsection{Spiking Neural Networks}
SNNs~\cite{Ponulak_Kasinski_2011, DBLP:reference/nc/Paugam-MoisyB12, DBLP:journals/nn/TavanaeiGKMM19} are a type of neural network model that aim to replicate the behavior of biological neurons. Unlike traditional DNNs that use continuous activation values, SNNs communicate through discrete electrical impulses called spikes. As shown in Figure~\ref{fig:SNN}b, these spikes encode the timing and intensity of neuron activations, allowing for more precise and efficient information processing~\cite{DBLP:conf/ijcnn/MarchisioNKHM020, DBLP:conf/iros/MarchisioPMM021, DBLP:conf/date/El-AllamiM0A21, DBLP:journals/neuromorphic/KimCP22}. SNNs are particularly suited for modeling dynamic and time-varying data, as they can capture the temporal aspects of input signals. This enables SNNs to excel in temporal pattern recognition, event-based processing, and real-time sensory processing~\cite{DBLP:series/sci/MeftahLCKB13, DBLP:conf/ijcnn/VialeMMMS21, DBLP:conf/ijcnn/CordoneMT22, DBLP:conf/iros/VialeMMM022}. {SNNs provide an efficient and brain-inspired computing paradigm for executing ML workloads. However, processing SNNs on traditional (Von Neumann) architectures demands high energy consumption and execution time. To overcome these issues, designers have developed specialized hardware platforms such as neuromorphic chips to execute SNNs in a fast and efficient manner. Compared to non-spiking DNNs, the communication between neurons in SNNs is discrete through spike trains, whereas DNNs have continuous activation values. The key advantage of SNNs is that computations are executed only in the presence of spikes. If the spikes are sparse in time, SNNs can save a large amount of energy compared to the non-spiking DNNs that process continuous values.} By emulating the spiking behavior of biological neurons, SNNs offer a promising avenue for understanding and replicating the computational capabilities of the human brain. Because conventional ML datasets typically lack any form of temporal encoding, an additional encoding step is necessary to introduce the required temporal dimension~\cite{DBLP:conf/ijcnn/MassaMM020}. In the case of SNNs, input spikes are treated as a sequence of tensors consisting of binary values~\cite{DBLP:journals/spm/IndiveriS19, DBLP:journals/corr/LeeDP16, DBLP:journals/corr/abs-1903-06379}.

\subsection{Homomorphic Encryption and Brakerski/Fan-Vercauteren scheme}
\begin{figure}[h]
    \centering
    \includegraphics[width=0.7\linewidth]{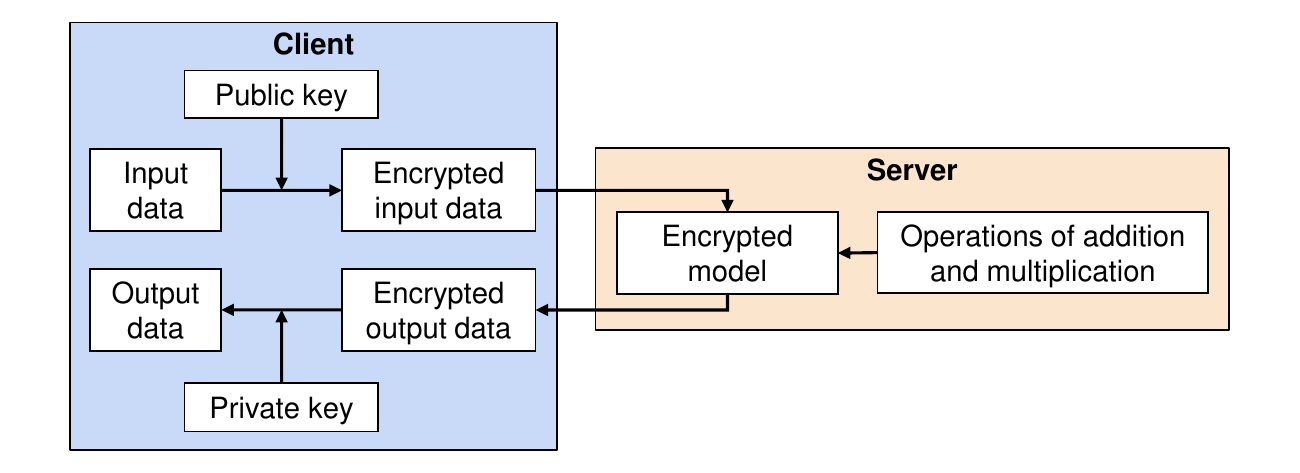}
    \caption{A fully homomorphic encryption (FHE) scheme.}
    \label{fig:encryption}
\end{figure}

HE is a cryptographic technique that allows computations on encrypted data without decryption~\cite{DBLP:conf/stoc/Gentry09, DBLP:journals/ejisec/OrlandiPB07}. A popular scheme used in HE is the BFV scheme~\cite{DBLP:journals/iacr/FanV12} (see Figure~\ref{fig:encryption}). This scheme leverages polynomial encoding to enable encrypted data manipulation. In this scheme, the client encrypts their sensitive input data using a public key provided by the server~\cite{DBLP:conf/asiacrypt/StehleSTX09, DBLP:conf/pkc/DamgardJ01}. The server computes the encrypted data using specialized algorithms that maintain the encryption. The encrypted results are then returned to the client, who can decrypt them using their private key to obtain the desired outputs. The BFV scheme supports addition and multiplication operations on encrypted variables, preserving the algebraic structures necessary for computation. By employing this scheme, sensitive data remain protected throughout the computation process, ensuring privacy and security~\mbox{\cite{Rivest1978ONDB, DBLP:journals/iacr/BosLLN13, DBLP:journals/iacr/ChabanneWMMP17, DBLP:journals/cim/FalcettaR22}.} HE comes in different variants, such as partially HE (PHE), somewhat HE (SHE), and fully HE (FHE), each offering different levels of computation capabilities on encrypted data~\cite{DBLP:journals/siamcomp/BrakerskiV14, DBLP:phd/us/Gentry09, DBLP:journals/iacr/BrakerskiGV11, DBLP:journals/iacr/FanV12, DBLP:conf/tcc/BonehGN05}.

{The BFV scheme is a type of FHE, which means that operations are fully encrypted, both on multiplications and additions. Consequently, there is no possibility of obtaining intermediate information during the process. To explain this concept more clearly, we can look at an example using an equation. In this case, we will apply homomorphic invariance only to addition, but FHE applies the same logic to multiplication as well. \linebreak Our basic equation is Equation~(\ref{eq:1}). Let us assume it undergoes a homomorphic transformation (encryption) represented as Equation~(\ref{eq:2}). Let us calculate the result by choosing random values for $x$ and $y$ (see Equation~(\ref{eq:3})). Calculating both sides of Equation (\ref{eq:1}), we obtain Equations~(\ref{eq:4}) and~(\ref{eq:5}). Applying the homomorphic transformation of Equation~(\ref{eq:2}), we obtain Equations~(\ref{eq:6}) and~(\ref{eq:7}).
We obtained the same result on both sides of the equation, despite the homomorphic transformation applied in the middle. This is what HE accomplishes. In the case of the BFV scheme and FHE in general, homomorphism applies to both additions and multiplications.}

\begin{equation}
    f(x+3y) = f(x) + f(3y)
    \label{eq:1}
\end{equation}

\begin{equation}
    f(z) = 5z
    \label{eq:2}
\end{equation}

\begin{equation}
    \begin{cases}
    x = 2; 
    y = -6
    \end{cases}
    \label{eq:3}
\end{equation}

\begin{equation}
    f(2+3 \cdot (-6)) = f(2) + f(3 \cdot (-6))
    \label{eq:4}
\end{equation}

\begin{equation}
    f(-16) = f(2) + f(-18)
    \label{eq:5}
\end{equation}

\begin{equation}
    -80 = 10 - 90
    \label{eq:6}
\end{equation}

\begin{equation}
    -80 = -80
    \label{eq:7}
\end{equation}

\section{\upshape\textbf{Proposed Encryption Framework}}
\label{sec:proposed_encryption_framewrok}

\begin{figure}[h]
    \centering
    \includegraphics[width=0.7\linewidth]{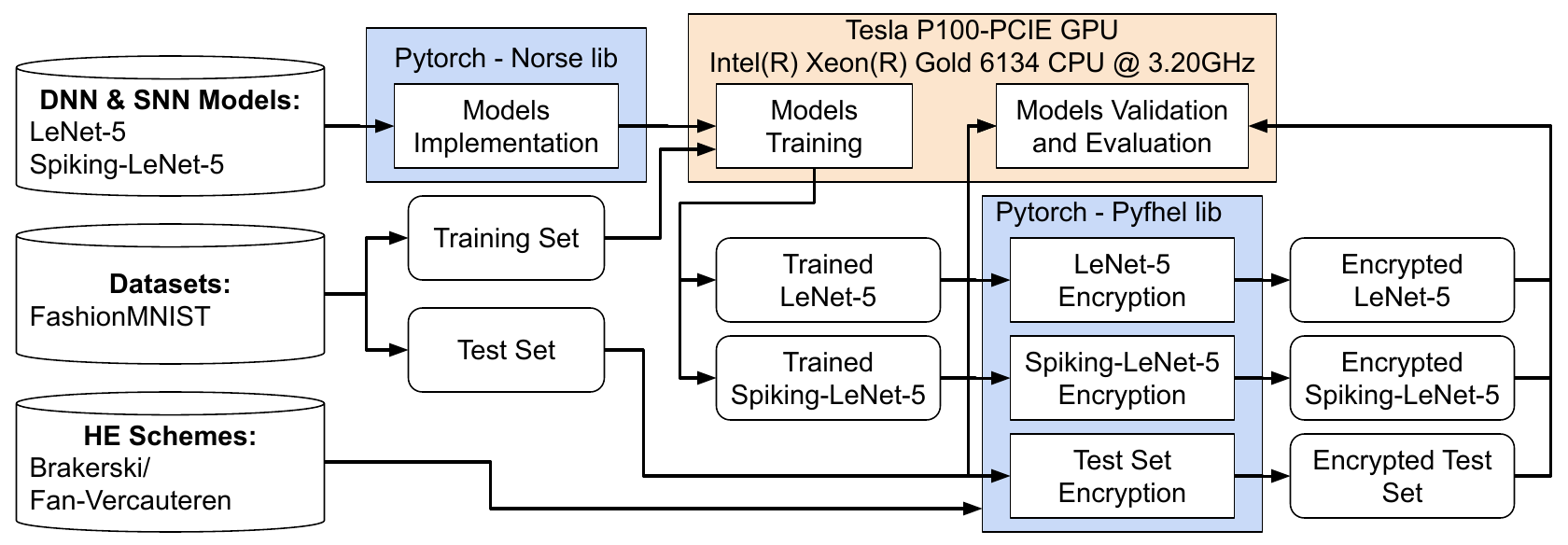}
    \caption{Our proposed encryption framework with the experimental setup.}
    \label{fig:flow_chart}
\end{figure}

In this work, (see Figure~\ref{fig:flow_chart}), we implement a LeNet-5 CNN~\cite{DBLP:journals/pieee/LeCunBBH98} and its equivalent SNN variant. For the dataset, we leveraged FashionMNIST~\cite{DBLP:journals/corr/abs-1708-07747} \mbox{(see Figure~\ref{fig:image_label_fashion})}, which is similar to MNIST~\cite{DBLP:journals/spm/Deng12} but consists of 10 classes of clothing items ({{note that we adopt the same test conditions as widely used by the SNN research community where the typical evaluation settings~\cite{Patino-Saucedo_2020NN_SNN_SpiNNaker} use the spiking LeNet and datasets such as MNIST and Fashion MNIST}}). \linebreak The hardware system used for conducting the experiments consisted of a Tesla P100-PCIE GPU, an Intel(R) Xeon(R) Gold 6134 CPU @ 3.20GHz, and 100 GB of RAM. We developed the code in Python, utilizing the PyTorch framework~\cite{DBLP:conf/nips/PaszkeGMLBCKLGA19}, the Pyfhel library for the encryption~\cite{DBLP:conf/ccs/IbarrondoV21}, and the Norse library to implement the SNN~\cite{pehle_2021}.

\subsection{FashionMNIST}
\begin{figure}[h]
    \centering
    \includegraphics[width=.7\linewidth]{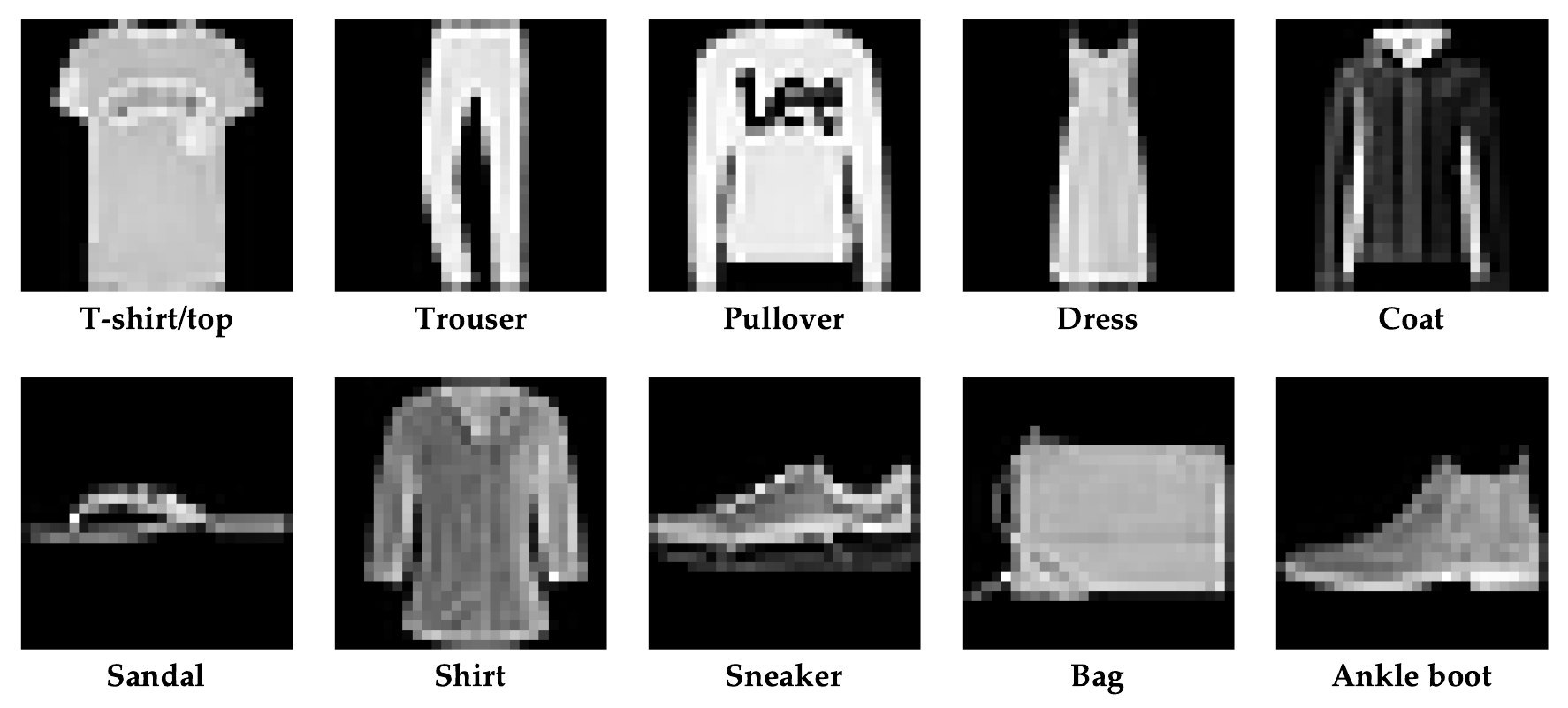}
    \caption{The FashionMNIST dataset consists of 10 classes of monochrome clothing items and is divided into 60,000 images for the training set and 10,000 images for the test set.}
    \label{fig:image_label_fashion}
\end{figure}

FashionMNIST~\cite{DBLP:journals/corr/abs-1708-07747} (see Figure~\ref{fig:image_label_fashion}) is a widely used dataset in computer vision and machine learning. It serves as a benchmark for image classification tasks and is a variation of the classic MNIST dataset. Instead of handwritten digits, FashionMNIST consists of grayscale images of various clothing items, such as shirts, dresses, shoes, and bags. It contains 60,000 training and 10,000 testing samples, each a 28 $\times$ 28 pixel image. The dataset offers a diverse range of clothing categories, making it suitable for evaluating algorithms and models for tasks such as image recognition, object detection, and fashion-related applications. FashionMNIST provides a challenging yet realistic dataset for researchers and practitioners to explore and develop innovative solutions in computer~vision.

\subsection{LeNet-5 and AlexNet}
\begin{figure}[h]
    \centering
    \includegraphics[width=1\linewidth]{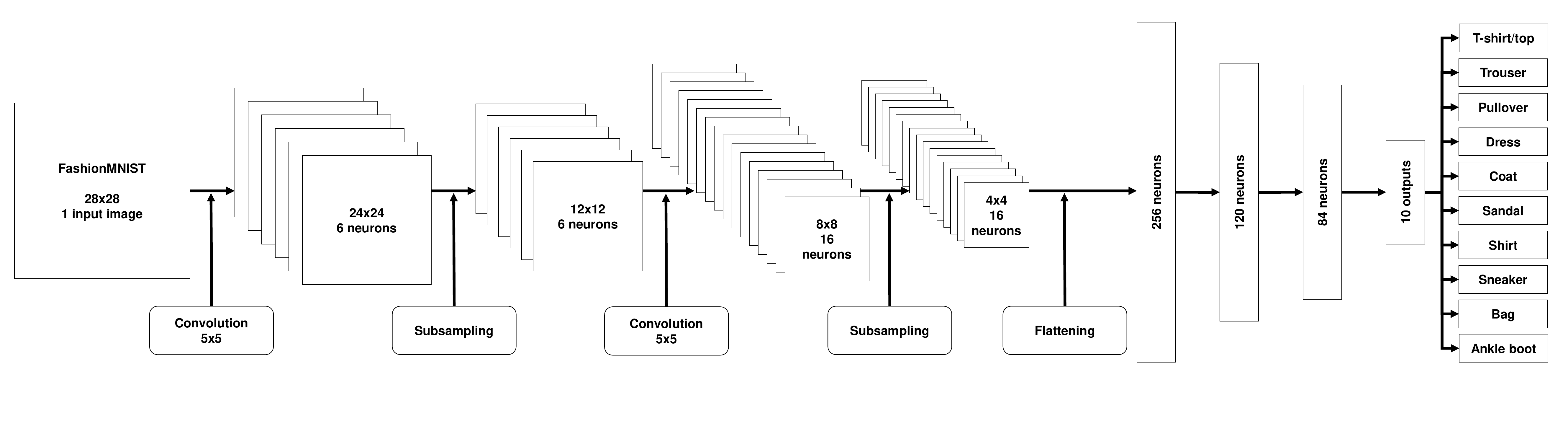}
    \caption{The LeNet-5 architecture, applied to the FashionMNIST dataset, used for the research.}
    \label{fig:lenet-5}
\end{figure}

LeNet-5 is a classic CNN architecture developed by Yann LeCun~\cite{DBLP:journals/pieee/LeCunBBH98}. It was explicitly designed for handwritten digit recognition and played a crucial role in the early advancements of deep learning. LeNet-5 is composed of convolutional, pooling, and fully connected layers (see Figure~\ref{fig:lenet-5}). The convolutional layers extract features from the input images using convolutional filters. The pooling layers reduce the dimensionality of the extracted features while preserving their essential information. Finally, the fully connected layers classify the features and produce the output predictions. LeNet-5 revolutionized the field of computer vision by demonstrating the effectiveness of CNNs for image classification tasks. Since then, it has served as a foundational model for developing more advanced CNN architectures and has found applications in various domains, including character recognition, object detection, and facial recognition.

{AlexNet~\cite{DBLP:journals/cacm/KrizhevskySH17} is a nine-layer DNN composed of six convolutional layers and three fully-connected layers. It represents the reference model for deep CNNs, where stacking several layers resulted in significant performance improvements compared to shallower CNNs. \linebreak A sequence of several convolutional layers can learn high-level features from the inputs that are used by fully connected layers to generate the output predictions.}

\subsection{Spiking-LeNet-5 and Norse, Spiking-AlexNet and Norse}
Spiking-LeNet-5~\cite{DBLP:journals/corr/abs-1802-02627, DBLP:journals/tnn/Izhikevich03, Roy2019TowardsSM, DBLP:conf/eccv/HanR20} is an extension of the LeNet-5 CNN architecture that incorporates the principles of SNNs~\cite{DBLP:journals/neco/ZenkeG18}. It is specifically designed to process temporal data encoded as spike trains, mimicking the behavior of biological neurons. Unlike the traditional LeNet-5, which operates on static input values, Spiking-LeNet-5 receives input spikes as a sequence of binary tensors. It utilizes specialized spiking neuron models, such as the leaky integrate-and-fire (LIF) neuron, to simulate the firing behavior of biological neurons~\cite{DBLP:journals/neco/PonulakK10}. The temporal dimension introduced by spike encoding allows Spiking-LeNet-5 to capture the dynamics and temporal dependencies present in the data. \linebreak This enables the network to learn and recognize patterns over time, making it suitable for tasks involving temporal data, such as event-based vision, audio processing, and other time-dependent applications. Spiking-LeNet-5 combines the power of traditional CNNs with the temporal processing capabilities of SNNs, opening up new possibilities for advanced SNN architectures. {Similarly, Spiking-AlexNet~\cite{Guo_2023Frontiers_TrainingSNN} extends AlexNet by incorporating the principles of SNNs, such as spike trains and LIF neurons.}

The LIF parameters~\cite{DBLP:journals/iacr/ChenLP17} in Norse are specific settings that define the behavior of LIF neurons in SNNs. These parameters include:
\begin{itemize}
\item \textbf{$\tau_{syn}^{-1}$}---represents the inverse of the synaptic time constant. It determines the rate at which the synaptic input decays over time;
\item \textbf{$\tau_{mem}^{-1}$}---represents the inverse of the membrane time constant. This parameter influences the rate at which the neuron's membrane potential decays without input;
\item \textbf{$v_{leak}$}---specifies the leak potential of the neuron. It is the resting potential of the neuron's membrane when there is no synaptic input or other stimuli;
\item \textbf{$v_{th}$}---defines the threshold potential of the neuron. The neuron generates an action potential when the membrane potential reaches or exceeds this threshold;
\item \textbf{$v_{reset}$}---represents the reset potential of the neuron. After firing an action potential, the membrane potential is reset to this value.
\end{itemize}

These parameters play a crucial role in shaping the dynamics of the LIF neuron in the SNN. They determine how the neuron integrates and responds to incoming synaptic input and when it generates an action potential. The specific values of these parameters can be adjusted to achieve desired behavior and control the firing rate and responsiveness of the neuron within the network.

SNNs also require an encoder because they operate on temporal data represented as spikes. Because most ML datasets do not include any temporal encoding, it is necessary to add an encoding step to provide the required temporal dimension. The encoder transforms the input data into sequences of spikes, which are then processed by the SNN as tensors containing binary values. The constant-current LIF encoder is an encoding method used in the Norse library to transform input data into sparse spikes. This encoding technique converts the constant input current into constant voltage spikes. During a specified time interval, known as $seq_{length}$, spikes are simulated based on the input current. This encoding allows Norse to operate on sparse input data in a sequence of binary tensors, which the SNN can efficiently process.

\subsection{HE parameters and Pyfhel}
\label{subsec:HE_parameters}

The HE process, implemented in the Pyfhel library, allows computations on encrypted data without decryption, ensuring data privacy and security~\cite{DBLP:journals/corr/PapernotMSW16, DBLP:conf/focs/Yao82b}. Pyfhel is built on the BFV scheme, a fully HE scheme.

The encryption process in the BFV scheme involves transforming plaintext data into ciphertext using a public key~\cite{DBLP:conf/eurocrypt/Paillier99}. The computations can be directly conducted on the ciphertext, preserving the confidentiality of the underlying plaintext~\cite{DBLP:conf/ijcnn/DisabatoFMR20}. The BFV scheme supports various mathematical operations on encrypted data, such as addition and multiplication. These operations can be performed on ciphertexts without decryption, enabling computations on sensitive data while maintaining its privacy~\cite{DBLP:conf/icml/Gilad-BachrachD16}.

The BFV scheme relies on three key parameters:
\begin{itemize}
\item \textbf{$m$}---represents the polynomial modulus degree, influencing the encryption scheme’s computational capabilities and security level; 
\item \textbf{$t$}---denotes the plaintext modulus and determines the size and precision of the encrypted plaintext values; 
\item \textbf{$q$}---represents the ciphertext modulus, determining the size of the encrypted ciphertext values and affecting the security and computational performance of the encryption scheme. 
\end{itemize}

A balance between security and computational efficiency in HE computations can be achieved by selecting appropriate values for these parameters. Pyfhel provides a convenient interface to work with the BFV scheme, allowing for data encryption, computation, and decryption while maintaining privacy and confidentiality.

Another critical parameter is the noise budget (NB), which refers to the maximum amount of noise or error that can be introduced during the encryption and computation process without affecting the correctness of the results. When performing computations on encrypted data, operations such as additions and multiplications can accumulate noise, deleting the decrypted results' accuracy. The NB represents a limit on how much noise can be tolerated before the decrypted results become unreliable. The NB needs to be carefully managed and monitored throughout the computation process to ensure the security and correctness of the encrypted computations.

\section{\upshape\textbf{Results and Discussion}}
\label{sec:results}

The experiments are divided into several parts to obtain accurate results:
\begin{itemize}
\item Training of the LeNet-5, AlexNet, Spiking-LeNet-5, and Spiking-AlexNet models on the training set of the FashionMNIST dataset;
\item Validating the models on the test set of the same dataset;
\item Creating encrypted models based on the previously trained models~\cite{DBLP:conf/aaai/KimVP22};
\item Encrypting the test set;
\item Evaluating the encrypted images on the encrypted LeNet-5, AlexNet, Spiking-LeNet-5, and Spiking-AlexNet models.
\end{itemize}

\subsection{Training phase}
For the training phase, optimal parameters were set to increase accuracy. The best learning rate was found using the learning rate finder technique~\cite{DBLP:conf/wacv/Smith17}, whereas the number of epochs was chosen based on early stopping to prevent overfitting~\cite{DBLP:conf/icml/RiceWK20}.
Table~\ref{tab_1} reports all the parameters chosen for the training phase.

\begin{table}[t] 
\centering
\caption{Training phase parameters.\label{tab_1}}
\begin{tabular}{ccccc}
\toprule
\textbf{Parameters}	                            & \textbf{LeNet-5}    & \textbf{Spiking-LeNet-5} & \textbf{AlexNet}    & \textbf{Spiking-AlexNet}\\
\midrule
Learning Rate		                            & 0.001               & 0.001 & 0.0001               & 0.0001\\
Epochs		                                    & 20	              & 20 & 20	              & 20\\
Optimizer~\cite{DBLP:journals/corr/KingmaB14}   & Adam                & Adam & Adam                & Adam\\
Loss~\cite{DBLP:journals/corr/JanochaC17}       & Cross Entropy       & Negative Log-Likelihood & Cross Entropy       & Negative Log-Likelihood\\ 
$seq_{length}$                                  & -                   & 30 & -                   & 30\\
$\tau_{syn}^{-1}$                               & -                   & 200 & -                   & 200\\
$\tau_{mem}^{-1}$                               & -                   & 100 & -                   & 100\\
$v_{leak}$                                      & -                   & 0 & -                   & 0\\
$v_{th}$                                        & -                   & 0.5 & -                   & 0.5\\
$v_{reset}$                                     & -                   & 0 & -                   & 0\\
Encoder                                         & -                   & Constant Current LIF & -                   & Constant Current LIF\\
\bottomrule
\end{tabular}
\end{table}

\begin{figure}[h]
    \centering
    \includegraphics[width=0.7\linewidth]{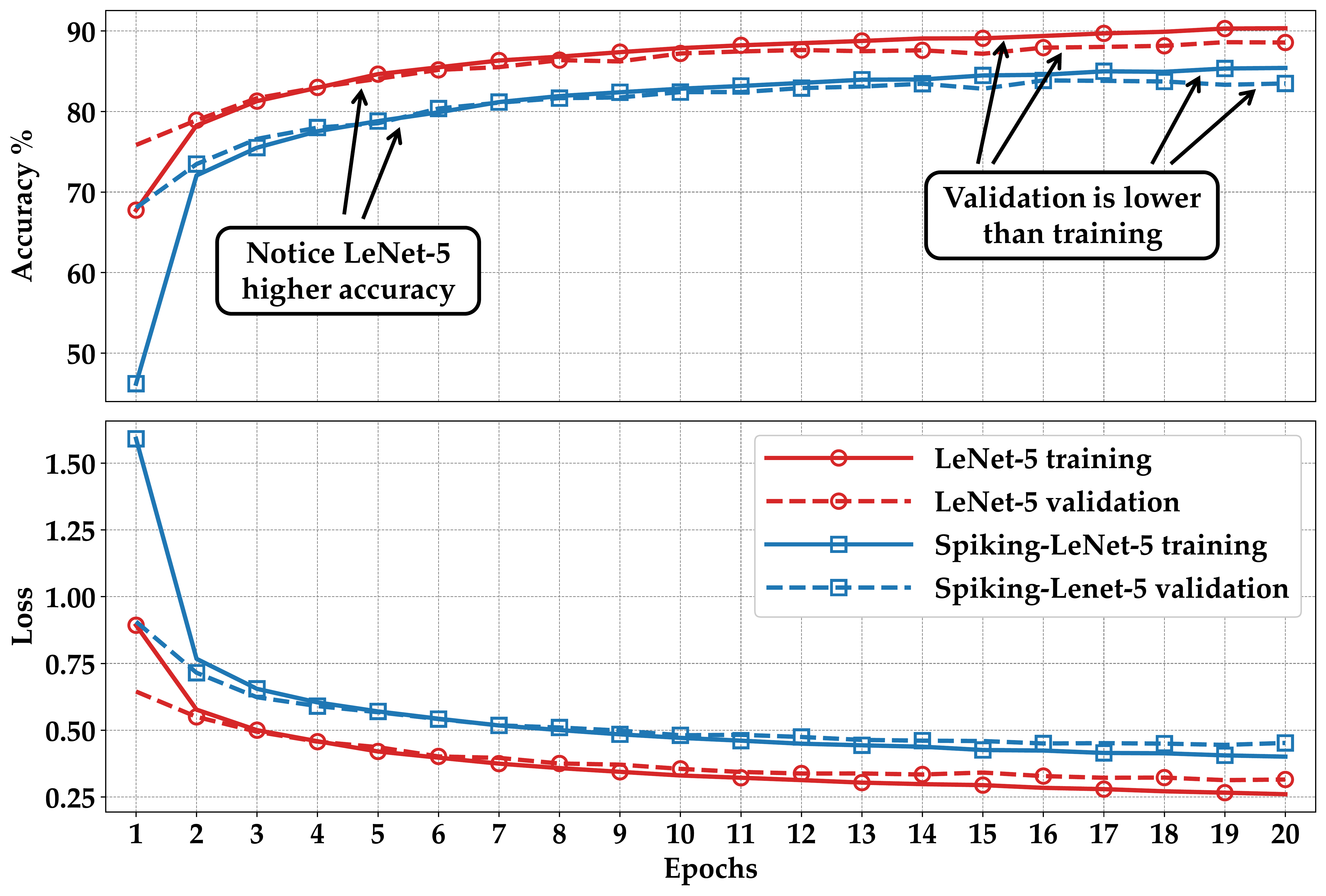}
    \caption{Accuracy and loss during training and validation of LeNet-5 and Spiking-LeNet-5 for the FashionMNIST dataset. The figure shows accuracy and loss values across different training~epochs.}
    \label{fig:line_accuracy_loss_training_fashion}
\end{figure}

Figure~\ref{fig:line_accuracy_loss_training_fashion} shows the accuracy and loss during training, comparing the LeNet-5 CNN with Spiking-LeNet-5 and their respective validation values at each epoch.

\unskip

Note that Spiking-LeNet-5 has slightly lower accuracy than (non-spiking) LeNet-5 due to the intrinsic complexity of the model itself, and its computational time is, on average, equal to that of LeNet-5 multiplied by the value of the $seq_{length}$.

\subsection{Encryption}
It is necessary to determine the three fundamental parameters that define a BFV HE scheme to proceed with image encryption: $m$, $t$, and $q$.

The $m$ parameter is chosen as a power of two and is directly proportional to the amount of NB. Values that are too small would be insecure, whereas values that are too large would make the computation too complex. Generally, $m$ is never less than 1024, and in our specific case, we observe that values of 2048 or higher do not influence the results but incur in exponentially longer computation time. For these reasons, we chose to keep the parameter $m$ fixed at 1024.

The $t$ parameter can also vary, and low values do not allow for proper encryption, whereas excessively high values degrade the result due to computational complexity. In our case, we evaluated the results over values ranging from 10 to 5000.

The $q$ parameter is closely related to the $m$ parameter in determining the NB. Hence, it is automatically calculated by the Pyfhel library to achieve proper encryption.

{With the hardware at our disposal (Tesla P100-PCIE GPU, Intel(R) Xeon(R) Gold 6134 @ 3.20GHz CPU, and 100 GB of RAM), it took approximately 30 s to encrypt each image and an additional 30 s to evaluate encrypted LeNet-5. However, for evaluation on encrypted Spiking-LeNet-5, it took around 15 min due to the $seq_{length}$ parameter equal to 30. For a clearer visualization, Table~\ref{tab_2} shows a comparison of the computation times for each image along with estimates for other models: AlexNet~\cite{DBLP:journals/cacm/KrizhevskySH17}, VGG-16~\cite{DBLP:journals/corr/SimonyanZ14a}, and ResNet-50~\cite{DBLP:conf/cvpr/HeZRS16}. These long execution times are aligned with the recent trend in the community that demands to build specialized accelerators for HE. A popular example is represented by the data protection in virtual environments (DPRIVE) challenge, used by DARPA to sponsor organizations that pursue R\&D of HE hardware~\cite{Cammarota_2022CCSW_Heracles, Badawi_2022CCS_OpenFHE, Cousins_2023arxiv_TREBUCHET}.}

\begin{table}[t] 
\centering
\caption{Execution time for each image reported in seconds for each model. The total encrypted execution is broken down into encryption and processing time of encrypted data. The long processing times of encrypted data are due to the complexity of the encrypted computations.}
\label{tab_2}
\begin{tabular}{ccccccccc}
\toprule
\textbf{Time (Seconds)}	 & \textbf{LeNet-5} & \textbf{S-LeNet-5}
& \textbf{AlexNet} & \textbf{S-AlexNet}
& \textbf{VGG-16} & \textbf{S-VGG-16}
& \textbf{ResNet-50} & \textbf{S-ResNet-50}\\
\midrule
Normal execution (unencrypted) & 0.03 & 1 & 30    & 1000   & 70    & 2300    & 10    & 300\\
Encrypted execution               & 31   & 930 & 30,060 & 901,800 & 70,140 & 2,104,200 & 10,020 & 300,600\\
\midrule
Encryption                        & 1    & 30  & 60    & 1800   & 140   & 4200    & 20    & 600\\
Processing time of encrypted data & 30   & 900 & 30,000 & 900,000 & 70,000 & 2,100,000 & 10,000 & 300,000\\
\bottomrule
\end{tabular}
\end{table}

\subsection{Evaluation}
\begin{figure}[h]
    \centering
    \includegraphics[width=0.7\linewidth]{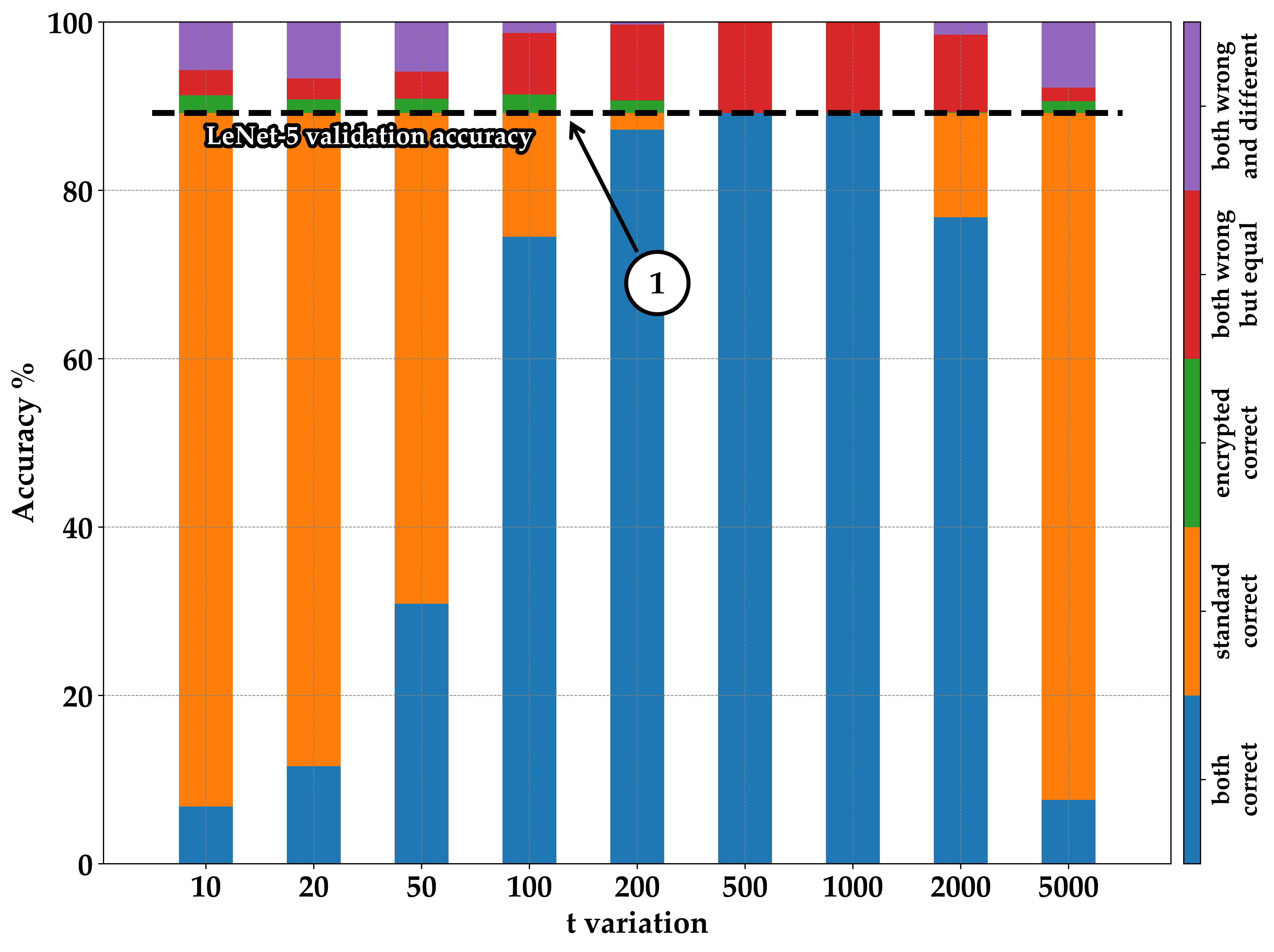}
    \caption{FashionMNIST accuracy on LeNet-5 for $t$ variation.}
    \label{fig:stackedbar_accuracy_t_variation_normal_fashion}
\end{figure}

\begin{figure}[h]
    \centering
    \includegraphics[width=0.7\linewidth]{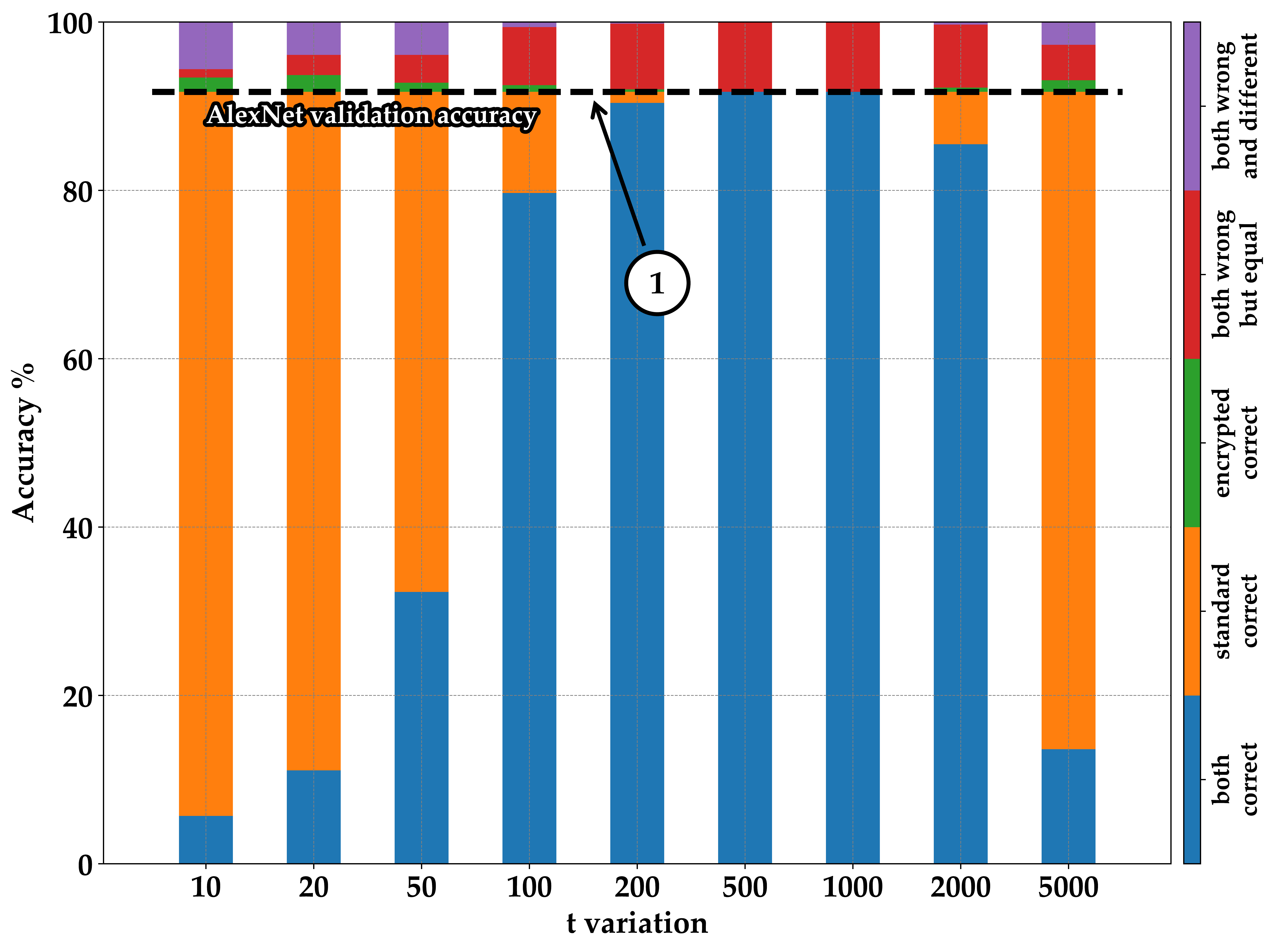}
    \caption{FashionMNIST accuracy on AlexNet for $t$ variation.}
    \label{fig:stackedbar_accuracy_t_variation_normal_fashion_alexnet}
\end{figure}

\begin{figure}[h]
    \centering
    \includegraphics[width=0.7\linewidth]{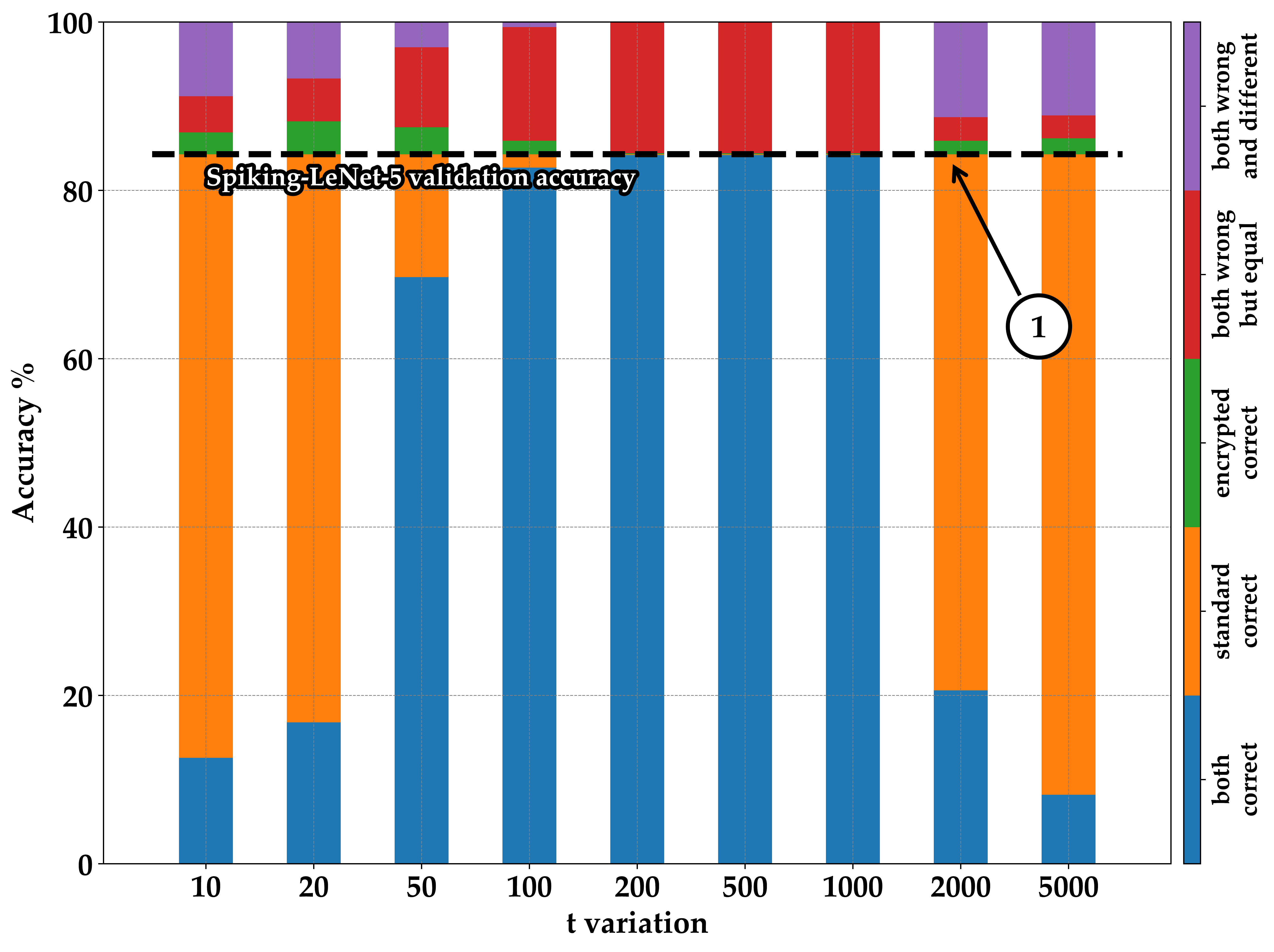}
    \caption{FashionMNIST accuracy on Spiking-LeNet-5 for $t$ variation.}
    \label{fig:stackedbar_accuracy_t_variation_spiking_fashion}
\end{figure}

\begin{figure}[h]
    \centering
    \includegraphics[width=0.7\linewidth]{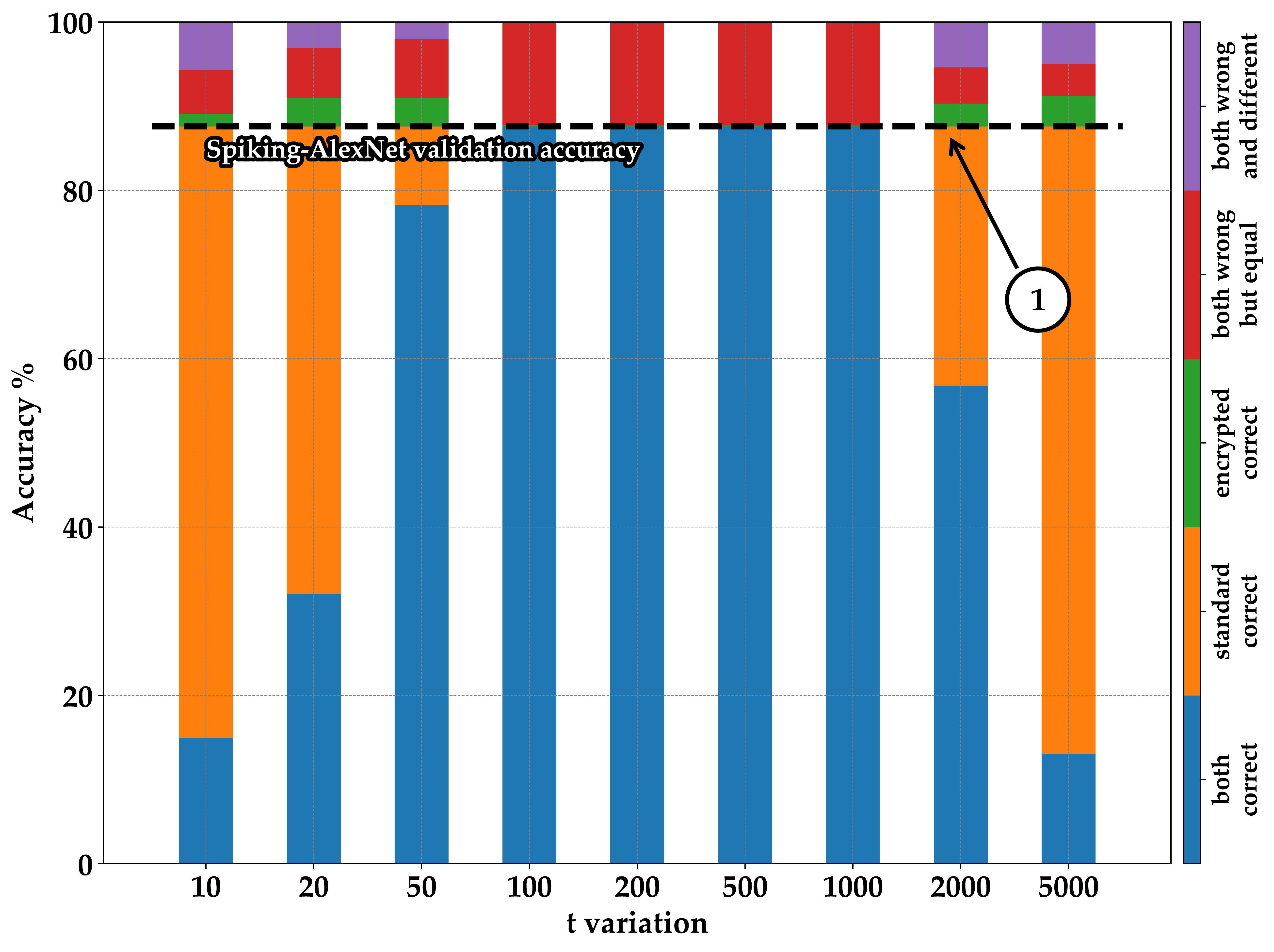}
    \caption{FashionMNIST accuracy on Spiking-AlexNet for $t$ variation.}
    \label{fig:stackedbar_accuracy_t_variation_spiking_fashion_alexnet}
\end{figure}

{In \Cref{fig:stackedbar_accuracy_t_variation_normal_fashion,fig:stackedbar_accuracy_t_variation_normal_fashion_alexnet,fig:stackedbar_accuracy_t_variation_spiking_fashion,fig:stackedbar_accuracy_t_variation_spiking_fashion_alexnet}}, we 
 can observe the results of encryption compared to the standard ones, along with the correct labels as the parameter $t$ varies.

\unskip

\unskip

\unskip

The various parts of the bars in the figures are divided as follows:
\begin{itemize} 
\item \colorbox{tab_blue}{\textcolor{white}{\textbf{Blue---both correct:}}} indicates the number of images classified correctly in both the standard and encrypted executions;
\item \colorbox{tab_orange}{\textcolor{white}{\textbf{Orange---standard correct:}}} represents the case where images are classified correctly in the standard execution but not in the encrypted one. It can be observed that by summing the blue and orange columns, we always obtain the same result: the accuracy of validation during training (see pointer~\rpoint{1}---\Cref{fig:stackedbar_accuracy_t_variation_normal_fashion,fig:stackedbar_accuracy_t_variation_normal_fashion_alexnet,fig:stackedbar_accuracy_t_variation_spiking_fashion,fig:stackedbar_accuracy_t_variation_spiking_fashion_alexnet}); 
\item \colorbox{tab_green}{\textcolor{white}{\textbf{Green---encrypted correct:}}} represents images classified correctly in the encrypted case but not in the standard one. It can be noticed that the percentages are generally low; this is because the encrypted model mistakenly classified the images differently from the standard model, but by chance, it happened to choose the correct label. Therefore, this column part does not represent a valid statistical case but rather randomness;
\item \colorbox{tab_red}{\textcolor{white}{\textbf{Red---both wrong but equal:}}} indicates cases where the encrypted model was classified identically to the standard one but did not classify the correct label. This part is essential, as it shows the encrypted model working correctly by emulating the standard model, even though the classification is incorrect overall;
\item \colorbox{tab_purple}{\textcolor{white}{\textbf{Purple---both wrong and different:}}} shows cases where the encrypted model made mistakes by not producing the same result as the standard model, and the standard model also made mistakes by not classifying correctly.
\end{itemize}

\begin{figure}[h]
    \centering
    \includegraphics[width=0.7\linewidth]{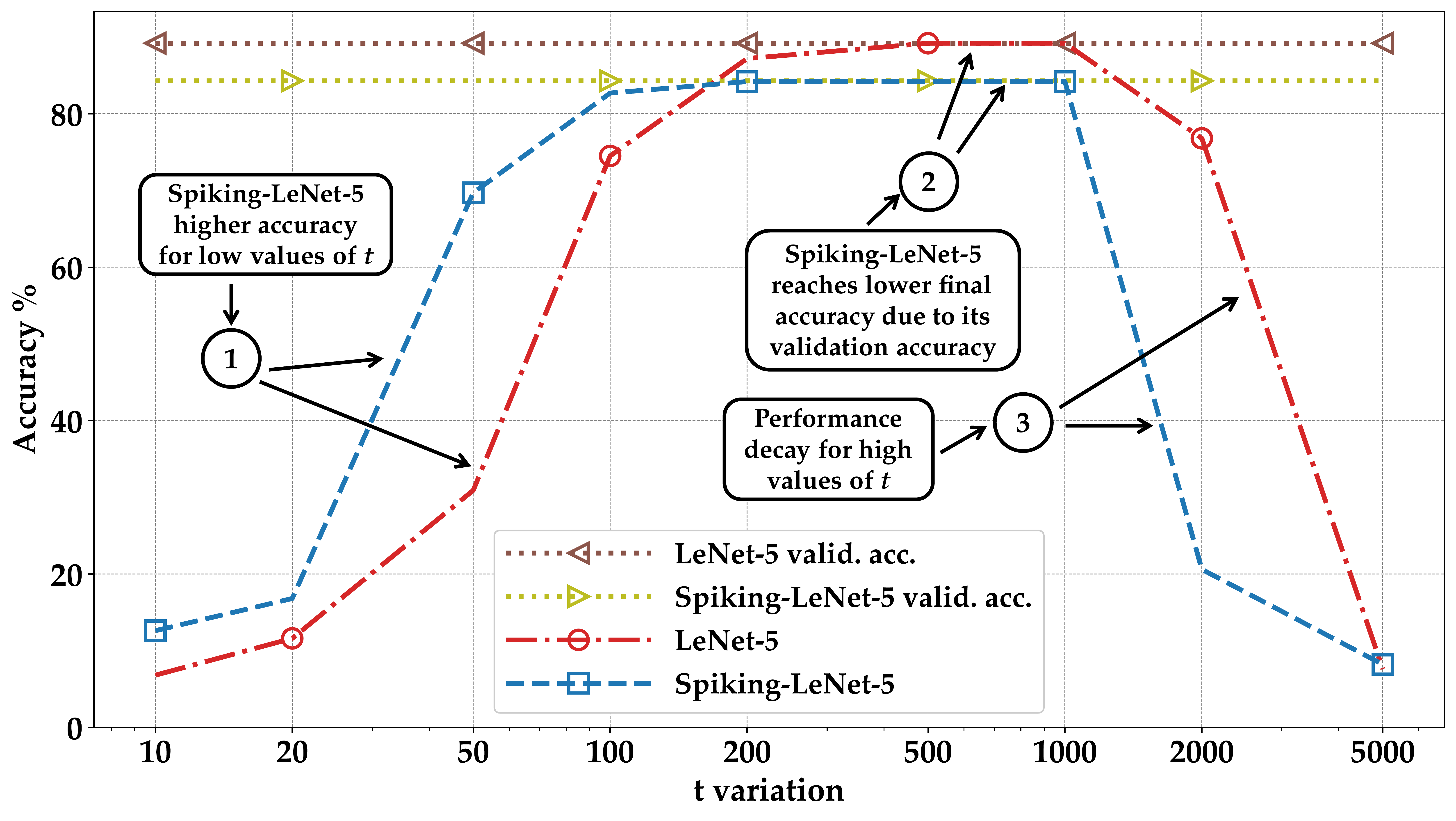}
    \caption{Comparison of FashionMNIST accuracy between LeNet-5 and Spiking-LeNet-5 for $t$ variations when both standard and encrypted versions classified correctly.}
    \label{fig:line_accuracy_total_fashion}
\end{figure}

\begin{figure}[h]
    \centering
    \includegraphics[width=0.7\linewidth]{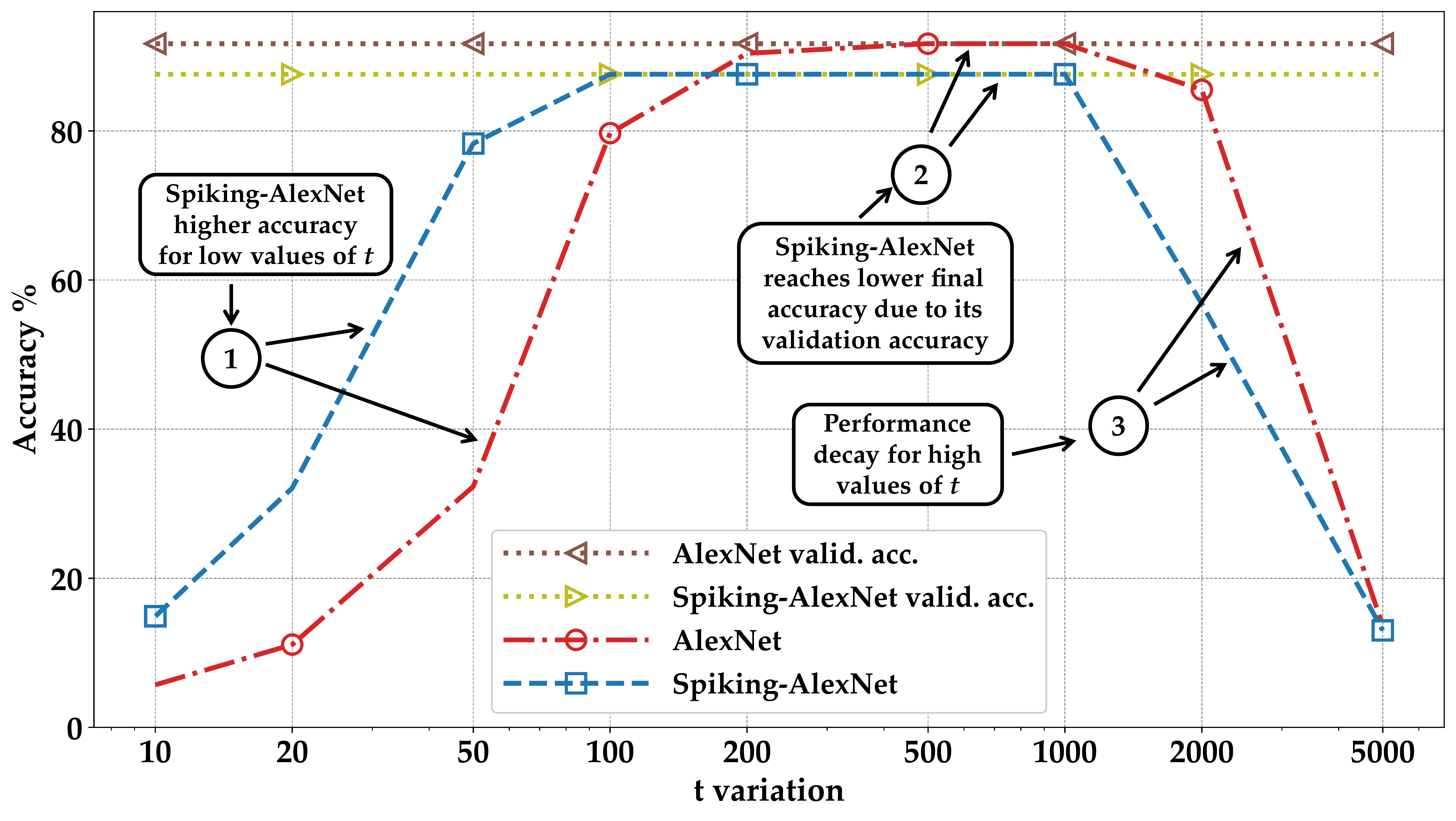}
    \caption{Comparison of FashionMNIST accuracy between AlexNet and Spiking-AlexNet for $t$ variations when both standard and encrypted versions classified correctly.}
    \label{fig:line_accuracy_total_fashion_alexnet}
\end{figure}

\begin{figure}[h]
    \centering
    \includegraphics[width=0.7\linewidth]{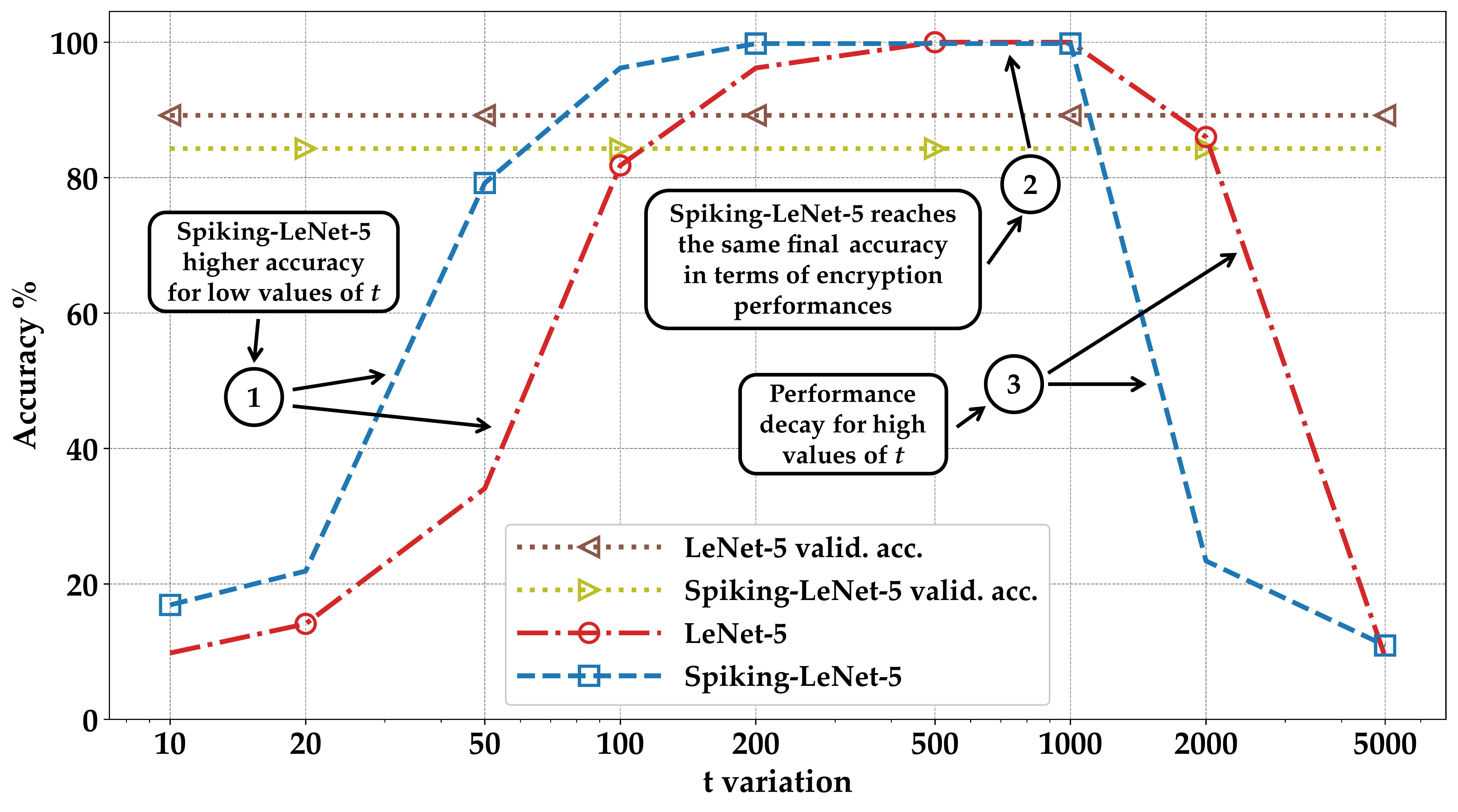}
    \caption{Comparison of FashionMNIST accuracy between LeNet-5 and Spiking-LeNet-5 for $t$ when the standard and encrypted versions coincide in both correct and incorrect classification.}
    \label{fig:line_accuracy_equal_fashion}
\end{figure}

\begin{figure}[h]
    \centering
    \includegraphics[width=0.7\linewidth]{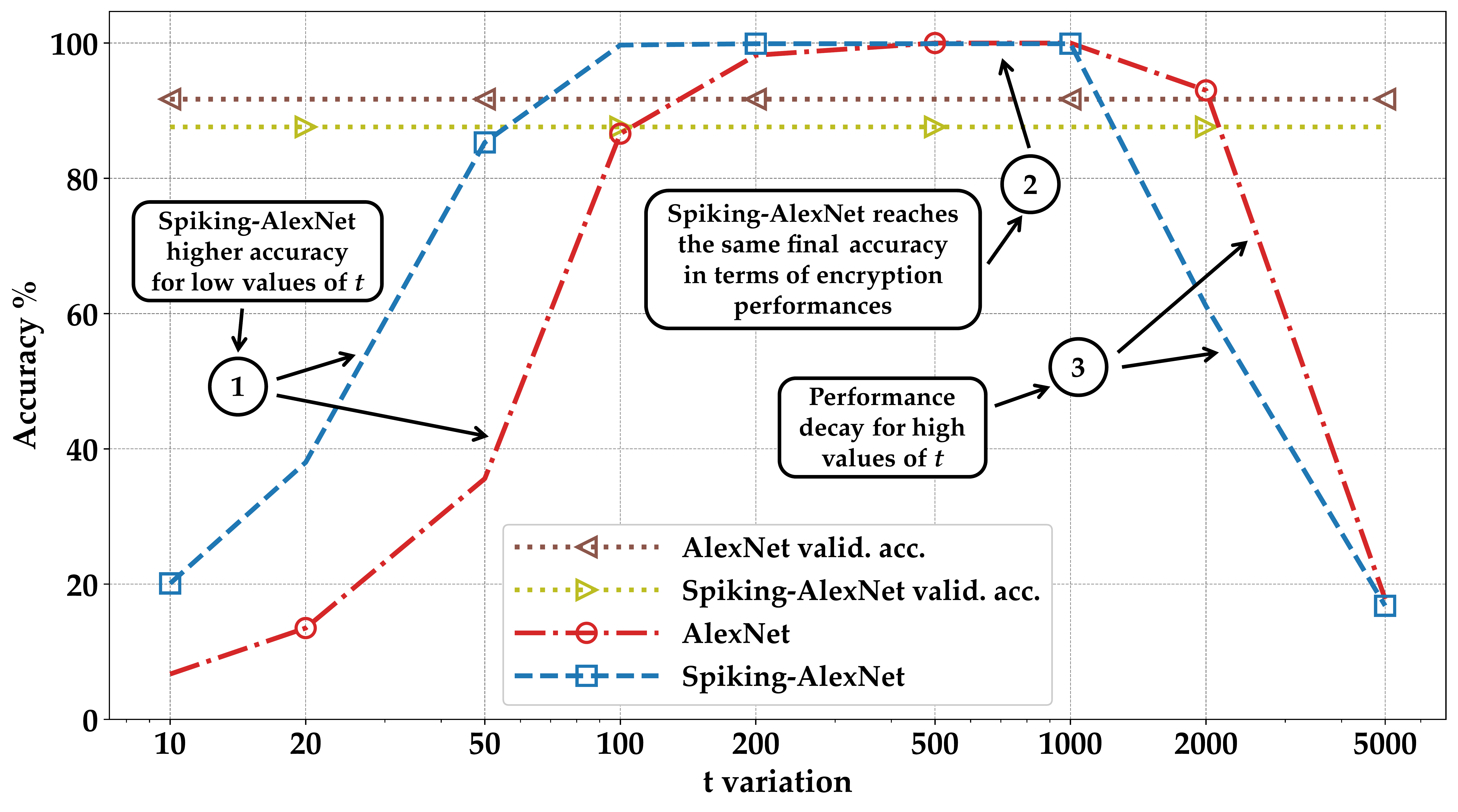}
    \caption{Comparison of FashionMNIST accuracy between AlexNet and Spiking-AlexNet for $t$ when the standard and encrypted versions coincide in both correct and incorrect classification.}
    \label{fig:line_accuracy_equal_fashion_alexnet}
\end{figure}

It can be noticed that for both low and high values of $t$, the results degrade rapidly. \linebreak For a better understanding, let us compare {LeNet-5 with Spiking-LeNet-5 by looking at Figures~\ref{fig:line_accuracy_total_fashion} and 
~\ref{fig:line_accuracy_equal_fashion}, and AlexNet with Spiking-AlexNet in Figures~\ref{fig:line_accuracy_total_fashion_alexnet} and~\ref{fig:line_accuracy_equal_fashion_alexnet}}, where the accuracies are graphically displayed as $t$ varies.

\unskip

\unskip

\unskip

{In Figures~\ref{fig:line_accuracy_total_fashion} and~\ref{fig:line_accuracy_total_fashion_alexnet}, we compared the LeNet-5, Spiking-LeNet-5, AlexNet, and Spiking-AlexNet} models in the case where both the standard and encrypted models were correct, representing the graphical representation of the blue parts of \Cref{fig:stackedbar_accuracy_t_variation_normal_fashion,fig:stackedbar_accuracy_t_variation_normal_fashion_alexnet,fig:stackedbar_accuracy_t_variation_spiking_fashion,fig:stackedbar_accuracy_t_variation_spiking_fashion_alexnet}. As can be seen, the Spiking-LeNet-5 version achieves acceptable levels of accuracy much earlier than LeNet-5, even with low values of $t$ (see pointer~\rpoint{1}---\Cref{fig:line_accuracy_total_fashion}). For instance, when \mbox{$t$ = 50}, Spiking-LeNet-5 achieves about 40\% higher accuracy than LeNet-5. However, the final accuracy of the Spiking-LeNet-5 model is slightly lower than that of LeNet-5 (see pointer~\rpoint{2}---\Cref{fig:line_accuracy_total_fashion}); this can be attributed to the fact that the Spiking-LeNet-5 model itself had lower validation accuracy compared to LeNet-5, as shown in Figure~\ref{fig:line_accuracy_loss_training_fashion}. {Similar observations can be derived by comparing AlexNet with Spiking-AlexNet. Spiking-AlexNet reaches higher accuracy than the AlexNet for low values of $t$ (see pointer~\rpoint{1}---\Cref{fig:line_accuracy_total_fashion_alexnet}), but for larger $t$, the accuracy of AlexNet is slightly higher than that of Spiking-AlexNet (see pointer~\rpoint{2}---\Cref{fig:line_accuracy_total_fashion_alexnet}).}

{On the contrary, in Figures~\ref{fig:line_accuracy_equal_fashion} and~\ref{fig:line_accuracy_equal_fashion_alexnet}, we compared the sums of the blue and red parts from \Cref{fig:stackedbar_accuracy_t_variation_normal_fashion,fig:stackedbar_accuracy_t_variation_normal_fashion_alexnet,fig:stackedbar_accuracy_t_variation_spiking_fashion,fig:stackedbar_accuracy_t_variation_spiking_fashion_alexnet}. In this manner, we can observe all the cases where the encrypted version produced the same result as the standard one, even if it was incorrect (see pointer~\rpoint{2}--- \Cref{fig:line_accuracy_equal_fashion,fig:line_accuracy_equal_fashion_alexnet}). From this graph, we can notice that the encrypted version of the Spiking-LeNet-5 model performs better than the encrypted LeNet-5, {and the encrypted Spiking-AlexNet performs better than the encrypted AlexNet. The SNNs achieve valid results with lower values of $t$ (see pointer~\rpoint{1}---\Cref{fig:line_accuracy_equal_fashion,fig:line_accuracy_equal_fashion_alexnet}) and higher overall accuracy.}
For excessively high values of $t$, {the results degrade for both the DNN and SNN models due to the increased computational complexity, which hinders the attainment of acceptable outputs (see pointer~\rpoint{3}--- \Cref{fig:line_accuracy_total_fashion,fig:line_accuracy_total_fashion_alexnet,fig:line_accuracy_equal_fashion,fig:line_accuracy_equal_fashion_alexnet}).}

\section{\upshape\textbf{Conclusions}}
\label{sec:conclusion}

In this work, we have demonstrated how SNNs can be a crucial factor in the development of future private and secure networks. Despite the increased time requirement, SNNs offer higher reliability, and further research can potentially reduce the time differences between DNNs and SNNs. The use of encryption systems such as HE is now more important than ever, considering the vast amount of data being exchanged worldwide. In this research, we have successfully shown how complete encryption systems can be applied to complex models, both CNNs and SNNs, ensuring correct final results without the possibility of decoding during the intermediate process, and how these results differ for CNNs and SNNs. {Our work represents the first proof-of-concept that demonstrates the applicability of HE schemes to SNNs. In future works, we plan to design acceleration techniques for encrypted SNNs and extend the experiment set with deeper networks.}

\section*{\upshape\textbf{Author Contributions}}

Conceptualization, F.N., R.C., A.M., M.M and M.S.; Methodology, F.N., R.C., A.M., M.M and M.S.; Software, F.N., R.C., and A.M.; Validation, F.N., R.C., and A.M.; Formal Analysis, F.N., R.C., and A.M.; Investigation, F.N., R.C., and A.M.; Resources, F.N., R.C., and A.M.; Data Curation, F.N., R.C., and A.M.; Writing – Original Draft Preparation, F.N., R.C., and A.M.; Writing – Review \& Editing, F.N., R.C., A.M., M.M and M.S.; Visualization, F.N., R.C., and A.M.; Supervision, F.N., A.M., M.M. and M.S.; Project Administration, M.M. and M.S.; Funding Acquisition, M.M. and M.S.

\section*{\upshape\textbf{Funding}}
This work has been supported in part by the Doctoral College Resilient Embedded Systems, which is run jointly by the TU Wien’s Faculty of Informatics and the UAS Technikum Wien.\linebreak This work was also supported in parts by the NYUAD Center for Cyber Security (CCS), funded by Tamkeen under the NYUAD Research Institute Award G1104, and the Center for Artificial Intelligence and Robotics (CAIR), funded by Tamkeen under the NYUAD Research Institute Award CG010.

\section*{\upshape\textbf{Data Availability Statement}}
Open-source framework: \href{https://github.com/farzadnikfam/SpyKing}{https://github.com/farzadnikfam/SpyKing}

\section*{\upshape\textbf{Conflicts of Interest}}
The funders had no role in the design of the study; in the collection, analyses, or interpretation of data; in the writing of the manuscript; or in the decision to publish the~results.

\section*{\upshape\textbf{Abbreviations}}
The following abbreviations are used in this manuscript:\\
\noindent 
\begin{tabular}{@{}ll}
ML & Machine Learning\\
DNNs & Deep Neural Networks\\
SNNs & Spiking Neural Networks\\
HE & Homomorphic Encryption\\
PHE & Partially Homomorphic Encryption\\
SHE & Somewhat Homomorphic Encryption\\
FHE & Fully Homomorphic Encryption\\
BFV & Brakerski/Fan-Vercauteren\\
CNNs & Convolutional Neural Networks\\
LIF & Leaky Integrate-and-Fire\\
NB & Noise Budget
\end{tabular}


\bibliographystyle{unsrt}

\bibliography{HE_SNN}

\vspace{0mm}

\section*{\upshape\textbf{Short Biography of Authors}}
\vskip -2\baselineskip plus -1fil

\begin{IEEEbiography}[{\includegraphics[width=1in,height=1.25in,clip,keepaspectratio]{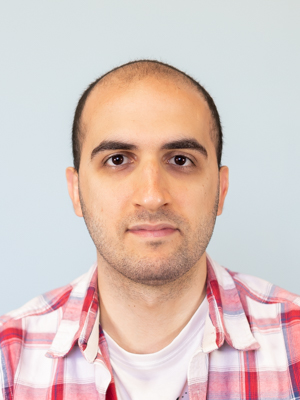}}]{Farzad Nikfam}
received the Bachelor’s degree in Mechanical Engineering, in March 2018, and the Master’s degree in Mechatronic Engineering, in March 2020 from Politecnico di Torino, Turin, Italy. Now he is pursuing a PhD in Electronics and Communication Engineering under the supervision of Prof. Maurizio Martina at Politecnico di Torino. His research activity is focused on software machine learning with emphasis on security and privacy problems.
\end{IEEEbiography}
\vskip -2\baselineskip plus -1fil

\begin{IEEEbiography}[{\includegraphics[width=1in,height=1.25in,clip,keepaspectratio]{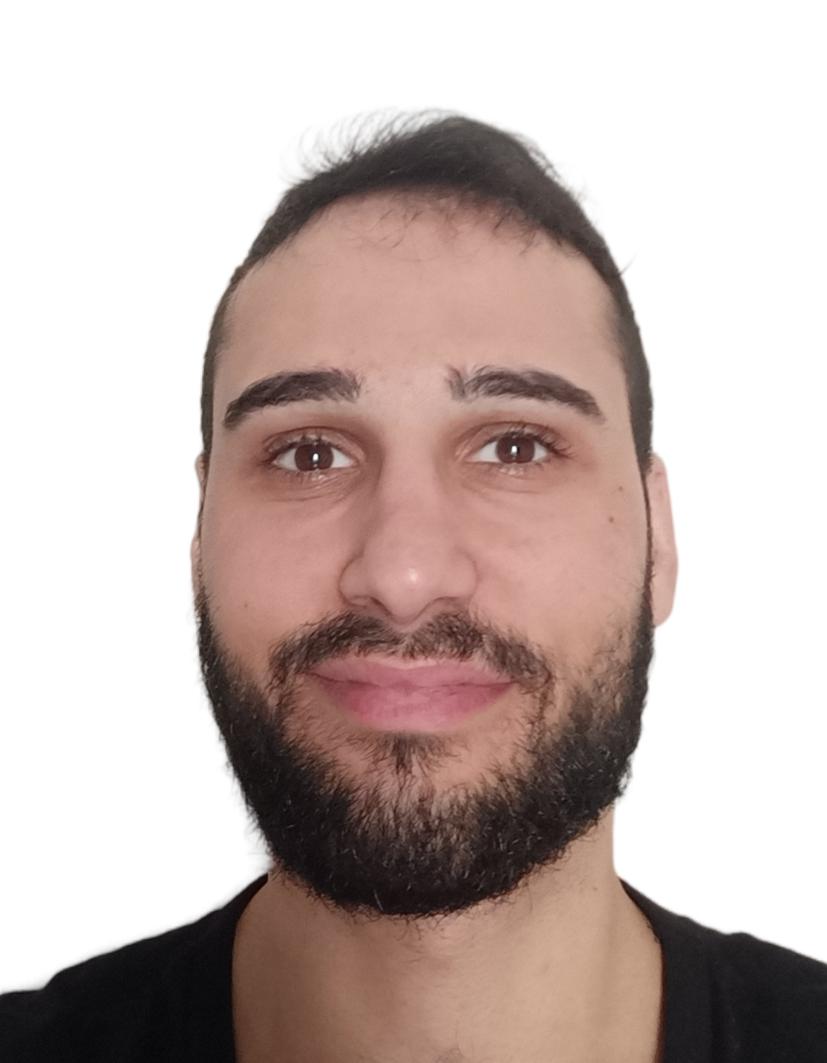}}]{Raffaele Casaburi}
is a junior engineer for STMicroelectronics in Catania. He enjoys contributing as a digital designer to the exciting technological advances the scientific world has to offer every day. He graduated from the University of Salerno in 2018 with a bachelor's degree in Electronic Engineering and earned a master’s degree in 2022 in Electronic Systems at Politecnico of Turin. Overall, his studies gave him the chance to develop efficient relationship skills and problem-solving capabilities in addition to his technical knowledge of digital architectures and implementation strategies. 
Moreover, the pre-graduate thesis experience let him explore the fascinating worlds of neural networks and encryption methodologies.
\end{IEEEbiography}
\vskip -2\baselineskip plus -1fil

\begin{IEEEbiography}[{\includegraphics[width=1in,height=1.25in,clip,keepaspectratio]{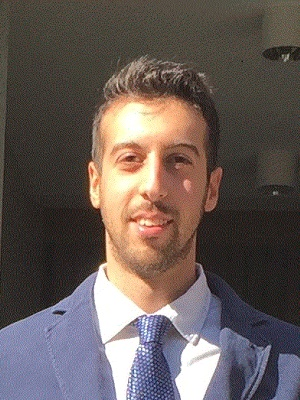}}]{Alberto Marchisio}
received his B.Sc. and M.Sc. degrees in Electronic Engineering from Politecnico di Torino, Turin, Italy, in October 2015 and April 2018, respectively. Currently, he is Ph.D. Student at Computer Architecture and Robust Energy-Efficient Technologies (CARE-Tech.) lab, Institute of Computer Engineering, Technische Universität Wien (TU Wien), Vienna, Austria, under the supervision of Prof. Dr. Muhammad Shafique. Since July 2023, he is with the Division of Engineering, New York University Abu Dhabi (NYUAD), United Arab Emirates. His main research interests include hardware and software optimizations for machine learning, brain-inspired computing, VLSI architecture design, emerging computing technologies, robust design, and approximate computing for energy efficiency. He (co-)authored 20+ papers in prestigious international conferences and journals. He received the honorable mention at the Italian National Finals of Maths Olympic Games in 2012, and the Richard Newton Young Fellow Award in 2019.
\end{IEEEbiography}
\vskip -2\baselineskip plus -1fil

\begin{IEEEbiography}[{\includegraphics[width=1in,height=1.25in,clip,keepaspectratio]{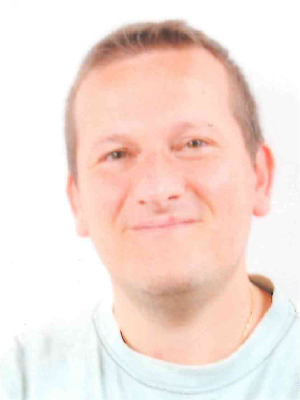}}]{Maurizio Martina}
received the M.S. and Ph.D. in electrical engineering from Politecnico di Torino, Italy, in 2000 and 2004, respectively. He is currently Full Professor with the VLSI-Lab group, Politecnico di Torino. His research interests include computer architecture and VLSI design of architectures for digital signal processing, video coding, communications, networking, artificial intelligence, machine learning and event-based processing. He edited one book and published 3 book chapters on VLSI architectures and digital circuits for video coding, wireless communications and error correcting codes. He has more than 100 scientific publications and is co-author of 2 patents. He has been an Associate Editor of IEEE Transactions on Circuits and Systems – I. He had been part of the organizing and technical committee of several international conferences, including BioCAS 2017, ICECS 2019, AICAS 2020. Currently, he is the counselor of the IEEE Student Branch at Politecnico di Torino and a professional member of IEEE HKN.
\end{IEEEbiography}
\vskip -2\baselineskip plus -1fil

\begin{IEEEbiography}[{\includegraphics[width=1in,height=1.25in,clip,keepaspectratio]{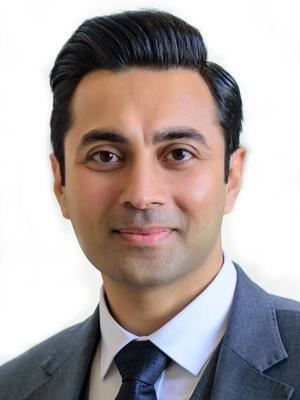}}]{Muhammad Shafique}
received the Ph.D. degree in computer science from the Karlsruhe Institute of Technology (KIT), Germany, in 2011. Afterwards, he established and led a highly recognized research group at KIT for several years as well as conducted impactful collaborative R\&D activities across the globe. In Oct.2016, he joined the Institute of Computer Engineering at the Faculty of Informatics, Technische Universität Wien (TU Wien), Vienna, Austria as a Full Professor of Computer Architecture and Robust, Energy-Efficient Technologies. Since Sep.2020, Dr. Shafique is with the New York University (NYU), where he is currently a Full Professor and the director of the eBrain Lab at the NYU-Abu Dhabi in UAE, and a Global Network Professor at the Tandon School of Engineering, NYU-New York City in USA. He is also a Co-PI/Investigator in multiple NYUAD Centers, including Center of Artificial Intelligence and Robotics (CAIR), Center of Cyber Security (CCS), Center for InTeractIng urban nEtworkS (CITIES), and Center for Quantum and Topological Systems (CQTS). His research interests are in AI \& machine learning hardware and system-level design, brain-inspired computing, autonomous systems, wearable healthcare, energy-efficient systems, robust computing, hardware security, emerging technologies, FPGAs, MPSoCs, and embedded systems. His research has a special focus on cross-layer analysis, modeling, design, and optimization of computing and memory systems. The researched technologies and tools are deployed in application use cases from Internet-of-Things (IoT), smart Cyber–Physical Systems (CPS), and ICT for Development (ICT4D) domains. Dr. Shafique has given several Keynotes, Invited Talks, and Tutorials, as well as organized many special sessions at premier venues. He has served as the PC Chair, General Chair, Track Chair, and PC member for several prestigious IEEE/ACM conferences. Dr. Shafique holds one U.S. patent has (co-)authored 6 Books, 15+ Book Chapters, 350+ papers in premier journals and conferences, and 100+ archive articles. He received the 2015 ACM/SIGDA Outstanding New Faculty Award, the AI 2000 Chip Technology Most Influential Scholar Award in 2020 and 2022, the ASPIRE AARE Research Excellence Award in 2021, six gold medals, and several best paper awards and nominations at prestigious conferences. He is a senior member of the IEEE and IEEE Signal Processing Society (SPS), and a member of the ACM, SIGARCH, SIGDA, SIGBED, and HIPEAC.
\end{IEEEbiography}
\vskip -2\baselineskip plus -1fil

\end{document}